**Exploring Machine Learning Models for Physical Dose Calculation in Carbon Ion Therapy Using Heterogeneous Imaging Data - A Proof of Concept Study.**

Miriam Schwarze[1,2,*], Hui Khee Looe[2], Björn Poppe[2], Pichaya Tappayuthpijarn[1], Leo Thomas[1] and Hans Rabus[1]

[1] Physikalisch-Technische Bundesanstalt, Abbestraße 2-12, 10587 Berlin, Germany
[2] Carl von Ossietzky Universität Oldenburg, Oldenburg, Germany
* corresponding author, miriam.schwarze@ptb.de

**Abstract**

*Background:* Accurate and fast dose calculation is essential for optimizing carbon ion therapy. Existing machine learning (ML) models have been developed for other radiotherapy modalities. They use patient data with uniform CT imaging properties.

*Purpose:* This study investigates the application of several ML models for physical dose calculation in carbon ion therapy and compares their ability to generalize to CT data with varying resolutions. Among the models examined is a Diffusion Model, which is tested for the first time for the calculation of physical dose distributions.

*Methods:* A dataset was generated using publicly available CT images of the head and neck region. Monoenergetic carbon ion beams were simulated at various initial energies using Geant4 simulation software. A U-Net architecture was developed for dose prediction based on distributions of material density in patients and of absorbed dose in water. It was trained as a Generative Adversarial Network (GAN) generator, a Diffusion Model noise estimator, and as a standalone network. Their performances were compared with two models from literature.

*Results:* All models produced dose distributions deviating by less than 2% from that obtained by a full Monte Carlo simulation, even for a patient not seen during training. Dose calculation time on a GPU was in the range of 3 ms to 15 s. The resource-efficient U-Net appears to perform comparably to the more computationally intensive GAN and Diffusion Model.

*Conclusion:* This study demonstrates that ML models can effectively balance accuracy and speed for physical dose calculation in carbon ion therapy. Using the computationally efficient U-Net can help conserve resources. The generalizability of the models to different CT image resolutions enables the use for different patients without extensive retraining.

## 1. Introduction

Carbon ion therapy is one of the most effective forms of radiation therapy, with particular benefits for deep-seated and radiation-resistant tumors. This effectiveness is attributed to the high linear energy transfer (LET) of carbon ions, which results in the induction of complex DNA damage and high radiobiological effectiveness (RBE). Moreover, the characteristic Bragg peak in the dose distribution of carbon ions enables precise dose delivery to the target volume. Carbon ions produce sharper Bragg peaks and have reduced lateral scattering compared to protons. However, nuclear reactions result in a more complex particle field.[1–4]

These properties are therapeutically advantageous solely with high precision treatment planning. In carbon-ion therapy, the RBE weighted dose is used as an optimization parameter in treatment planning, which is the product of RBE and physical dose. The RBE is determined using empirical models such as the Local Effect Model (LEM)[5–9], the Mixed-beam Model[10,11] or the Microdosimetric Kinetic Model (MKM)[12,13]. All models comprise a mixed functional equation, which depends on empirical parameters and the physical dose. Determining the distribution of the physical dose in the patient is therefore an essential component of treatment planning and is done using complex particle transport calculations.

Currently, analytical algorithms are typically used for this purpose[14–16]. While these algorithms provide a fast dose calculation, their accuracy is limited by various simplifications. For example, the patient geometry is often assumed to be homogeneous, the complexity of the mixed radiation field is not fully considered, and the three-dimensional (3D) spread of particle fluence is neglected[17]. Although Monte Carlo (MC) particle transport simulations are the most accurate approach to dose calculation, long computation times limit their clinical applications[18]. Therefore, numerous studies have focused on developing particle transport algorithms that combine the accuracy of MC simulations with short computation times.

Remarkable results for fast and accurate dose calculation were achieved using machine learning (ML) algorithms. These methods often use a two-step approach by converting computationally inexpensive precalculated dose or fluence distributions into dose distributions created by MC simulations. MC simulations with reduced statistical uncertainty[19–24] or analytical pencil beam algorithms[25,26] were employed as pre-calculation methods. While the transformation of the dose distributions by the neural network was completed within seconds, the time required for the pre-calculations was 16 minutes (5 min using high-performance computers) for carbon ions using noisy Monte Carlo simulation [20], and 2-5 min for protons using a pencil beam algorithms[25]. Comparable times are expected for carbon ion therapy[27].

Furthermore, pre-calculations were performed for the beam in homogeneous water, and the resulting dose distribution was subsequently transformed into the dose distribution of the inhomogeneous geometry (patient) based on the underlying density distribution[28]. As the precalculated dose distribution is independent of the patient geometry, it does not need to be recalculated for each case during dose calculation. Instead, a dataset covering all possible beam configurations can be generated in advance and the dose can be calculated in a single step by the generator. Thus, the time required to calculate a dose distribution becomes independent of the precalculations and was reported to be less than one second for x-ray microbeam therapy[28]. This was achieved using a generative adversarial network (GAN) [29,30] with a conditional U-Net[31] as the generator.

Similar computation times have been achieved with one-step approaches that do not require any precomputation. Mentzel et al.[32] proposed a modified U-Net for predicting dose distributions using the density distribution in geometry in the context of X-ray microbeam therapy, achieving computation times in the millisecond range. However, as the model does not incorporate information about the beam, this approach is not generalizable to different beam configurations. Pastor-Serrano et al.[33] introduced the Dose Transformer Algorithm (DoTA), a dose prediction model for protons based on CT images of the geometry and beam energy. The model architecture combines convolutional neural network (CNN) and Transformer[34] modules, treating the 3D geometry and dose matrices as a sequence of two-dimensional slices aligned along the beam direction. This method also achieves dose computation times in the millisecond range.

A comparison of DoTA, the GAN-based model[28] and the U-Net trained as a regression model, used therein as a generator, was presented in Mentzel et al.[35] for proton minibeams. Among these approaches, the U-Net models provide the fastest dose calculations, with the U-Net regression model achieving the highest accuracy.

These studies should be distinguished from approaches in which entire treatment plans are predicted; a recent review can be found in Kui et al.[36]. These methods learn the optimal dose distribution of an entire treatment plan within a given patient geometry, achieved by combining various beamlets. In contrast, the dose calculation algorithms discussed in this paper predict the physical dose distribution of a single beamlet with specific given properties in the patient geometry. However, the data structures are similar, making it worthwhile to examine the models used. Recent work demonstrates that Diffusion Models[37] achieve higher accuracy in predicting the optimal dose distribution in radiotherapy treatment planning than a GAN or U-Net model[38–42].

To date, the only published studies addressing physical dose calculation in carbon ion therapy using ML rely on extensive pre-calculations, either based on MC simulations or pencil beam algorithms [20,26]. An effective ML based algorithm for physical dose calculation should avoid the need for time-consuming precalculations during dose evaluation. As outlined above, two approaches that meet this criterion (for x-ray microbeams, proton minibeams and proton beams) have been published. Therefore, one key objective of this paper is to explore the transferability of these methods to carbon ion beams. This work is restricted to the calculation of the physical dose, as this is the main time-consuming factor in determining the optimization parameter in treatment planning for carbon ion therapy.

The first approach, proposed by Mentzel et al.[28,35,43], involves training ML models to transform a dose distribution in water into a dose distribution of the same beam in an inhomogeneous geometry, based on the density distribution. While the datasets used by Mentzel et al. consist of simplified head and rat phantoms composed of water, bone and air, the dataset used in this work encompasses the full variability of materials and densities found in real human heads. This extends the learning task, as a greater variability in materials with corresponding interaction behaviors must be learned. In addition, more interfaces are present within the geometry. Furthermore, the dose distribution produced by carbon ions differs significantly from the dose distribution of the x-ray microbeams and proton minibeams used by Mentzel et al. Therefore, a novel network architecture has been developed and, on the one hand, trained as a standalone U-Net and GAN, similar to the approach of Mentzel et al.[35]. On the other hand, it is trained as a noise estimator within a Diffusion Model. This represents the first application of a Diffusion Model for calculating a physical dose distribution. To assess the suitability of a Diffusion Model for predicting the physical dose more comprehensively, the architecture of DiffDP[38], which was developed for the prediction of an optimized dose distribution in conventional radiotherapy, is also tested for the same purpose.

The second approach, DoTA, was developed by Pastor-Serrano et al.[33] for proton therapy and trained on a dataset of real anatomical structures. The sharper Bragg peak of carbon ions, compared to protons, results in higher gradients in the dose distribution, which are more challenging to learn. Nevertheless, the complexity of the learning task is comparable, so DoTA is only retrained using carbon ion therapy data, without any modifications to the model architecture.

Both Mentzel et al.[28,35,43] and Pastor-Serrano et al.[33] use datasets with a constant voxel size. However, in clinical practice, CT images typically do not have the same resolution. Therefore, the third objective of this paper is to test and compare the generalizability of the models to different CT image resolutions. Consequently, our dataset includes CT images with varying resolutions.

## 2. Materials and Methods

### 2.1 Dataset and Preprocessing

The dataset comprises pairs of 3D density and dose distributions. It is based on eight randomly selected CT images of the head and neck region from the GLIS-RT (Glioma Image Segmentation for Radiotherapy) dataset of The Cancer Imaging Archive (TCIA)[44], hereafter referred to as patients. Dose calculations were performed using the MC simulation software Geant4 (version 11.0.2)[45–47]. Circular monoenergetic carbon ion beams with a diameter of 1 mm, initial energies ranging from 1250 to 3000 MeV (corresponding energy per mass 104.16 to 250 MeV/u) (in 125 MeV (10.42 MeV/u) steps), and a fixed start position outside the heads were simulated. The Bragg peak of the 3000 MeV carbon ion beam is approximately at the opposite side of the heads. The beam axis was from one side of the head to the other (Supplementary Figure S1 to S3). The source position was varied in 25 mm steps in the planes perpendicular to the beam axis. It should be noted that the beam properties used do not correspond to standard clinical practice. This study only examines the ability of the networks to learn the interaction mechanisms of the particles and the generalizability to different image resolutions. While the approaches could have been used with beams having energy and angular spread, it appears more straightforward to test the approaches with monoenergetic pencil beams. Testing the generalizability to different beam parameters is postponed to future work.

The resulting dataset contains 4885 samples. Each sample includes $10^7$ simulated primary carbon ions. A detailed description of the dataset is provided in Supplementary Figure S1 to Supplementary Figure S4, while the voxel size and number of the individual CT images are listed in Supplementary Table 1.

The simulation is based on Geant4's DICOM example[48]. This includes a DICOM reader that converts the CT images into a Geant4-compatible geometry and stores them as a density matrix. The voxelization of the CT image is preserved in this process. The DICOM example also includes functions for scoring the dose per voxel. The physics list was adapted to simulate carbon ions following the hadron therapy example[49]. The output is a list of voxel indices and their corresponding dose values, which are then transformed into a 3D dose matrix.

The density and dose matrices of all patients are resampled to the same voxel size of $(1.3 \times 1.3 \times 2.5)\,\text{mm}^3$ and cropped to a size of $(256 \times 16 \times 16)$ voxels, with the first dimension corresponding to the beam axis. Therefore the functions of SimpleITK[50–52] with nearest neighbor interpolation were used. In the planes perpendicular to the beam axis, the region of interest is chosen to center the source position.

In addition, a dose distribution in water is simulated for all carbon ion energies used, each including $10^7$ simulated primary carbon ions. To enable flexible changes in the dose distribution resolution without repeating the time-consuming simulations, the location of all energy transfer points and the corresponding energy deposits in a water phantom are scored. After the simulation, these values are transformed into a 3D dose distribution. The water dose matrices are given the same shape and resolution as the head dose matrices.

A total of 15 % (642 samples) of the samples are used as a validation dataset to monitor the training process and perform hyperparameter optimization. A further 15 % (644 samples) of the samples are retained during training and used later to test the models. The three datasets are randomly distributed across energies, patients and positions (Supplementary Figure S4). Furthermore, one randomly selected patient geometry is completely withheld from training and used to test the model´s generalizability to varying CT image and patient properties. To increase the training dataset, the matrices are rotated by 0 °, 90 °, 180 ° and 270 °, resulting in a fourfold increase in the original training dataset size (11 996 samples). All datasets are normalized to the global maximum (of all datasets and energies) using min-max normalization.

*2.2 Dose Prediction Models*

For the calculation of the physical dose, three models are examined, which are explained in more detail in the following sections: A U-Net, a GAN and a Diffusion Model. All models are based on the same conditions, as proposed by Mentzel et al.*[28,35,43]*: The density distribution of the patient geometry and the dose distribution of the beam in homogeneous water. The learning task for all models is therefore to transform this dose distribution from homogeneous water into the dose distribution of the same beam in the inhomogeneous patient.

This transformation is different in our application because, on the one hand, real patient data is used (Mentzel et al. used simplifying phantoms) and the shape of the dose distributions differs significantly (Mentzel et al. considered x-ray microbeams and proton minibeams). Therefore, a new U-Net architecture optimized for our dataset has been developed. The hyperparameters and model architecture were selected by iterative manual adjustment based on empirical intermediate results for the U-Net, using the lowest validation loss as the criterion for optimization. Only minor adjustments were made for the GAN generator and Diffusion Model noise estimator using the same criterion.

## 2.2.1 U-Net

A performance improvement was observed when the two input matrices were independently processed by separate encoders with identical structure, in contrast to the conventional method of input concatenation before a single encoder. This double encoder structure was proposed by Feng et al. in the DiffDP network[38]. While the general structure of DiffDP (double encoder, bottleneck and decoder) was retained, the block components, feature map fusion strategy, and the placement of skip connections were specifically optimized for our dataset. The best performance is achieved with five encoder levels, each containing a Residual Block (ResBlock)[53], Batch Normalization (BatchNorm)[54] layer, and downsampling convolutional layer. The latter has a kernel size of 3 and a stride of 2. Each ResBlock includes two convolutional, two BatchNorm, and two Rectified Linear Unit (ReLU) activation layers. As the network depth increases, the filter sizes increase to capture more complex features while the kernel sizes decrease to preserve small details, as detailed in Figure 1. After each level, the feature map of the density distribution is added to that of the dose distribution in water. At the final level, no downsampling occurs and the addition is replaced by a cross-attention module[34], where the density feature map acts as the key and value and the water feature map as the query vector.

The bottleneck comprises two additional ResBlocks separated by a BatchNorm layer. The decoder consists of four levels, each starting with a nearest neighbor upsampling layer. This is followed by two ResBlocks, a convolutional layer with a kernel size of 1, and another BatchNorm layer. Information from the encoder is passed to the decoder via skip connections, with two connections per level: One before and one after adding the feature maps in the encoder. The skip connection after the addition is concatenated after the second ResBlock, while the connection before the addition is inserted after the first ResBlock.

The U-Net was trained using the mean-squared error (MSE) between the predicted and the ground-truth dose distribution (obtained by the MC simulation) as a loss function.

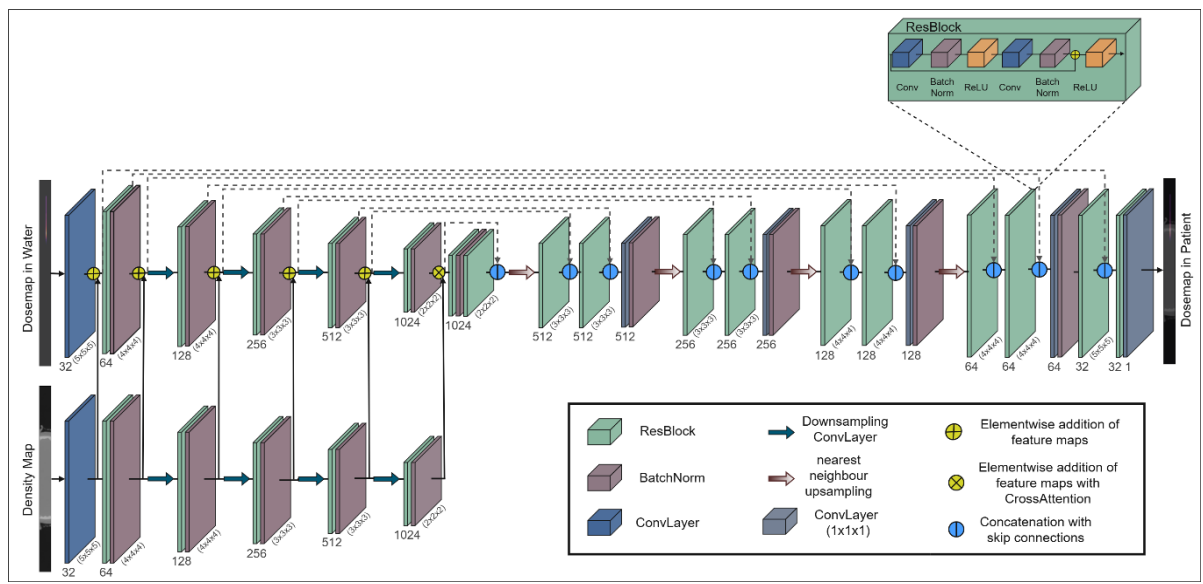

Figure 1: Architecture of the U-Net. The numbers below the boxes represent the respective filter size, and the numbers to the right of the boxes denote the kernel size.

### 2.2.2 GAN

The GAN comprises two competing neural networks: A generator and a discriminator. The generator is trained to produce synthetic dose distributions that match the distributions of the training data. The discriminator learns to distinguish between the real dose distributions and the synthetic ones created by the generator. The Wasserstein metric[30] with a gradient penalty term (gradient penalty coefficient $\lambda = 10$)[55] is used as a measure of the difference between the distributions. Both networks are trained simultaneously, with the generator aiming to deceive the discriminator by producing increasingly realistic dose distributions. Concurrently, the discriminator improves its ability to distinguish between synthetic and real distributions. One training step of the generator occurs after five training steps of the discriminator [29].

In the conditional GAN[56] used here, both models receive information about the patient geometry (density matrix) and beam parameters (dose distribution in water), allowing the dependency between the input information and dose distribution to be learned. Our generator uses the modified U-Net architecture described in Section 2.2.1. Our discriminator has the same encoder structure as the generator, as shown in Figure 2. The encoder is followed by four combinations, each consisting of a convolutional layer to reduce the matrix dimensions and a ReLU activation function. A flattening layer converts the 3D matrix into a one-dimensional vector. A linear layer outputs a real number representing the fidelity score of the entered dose distribution.

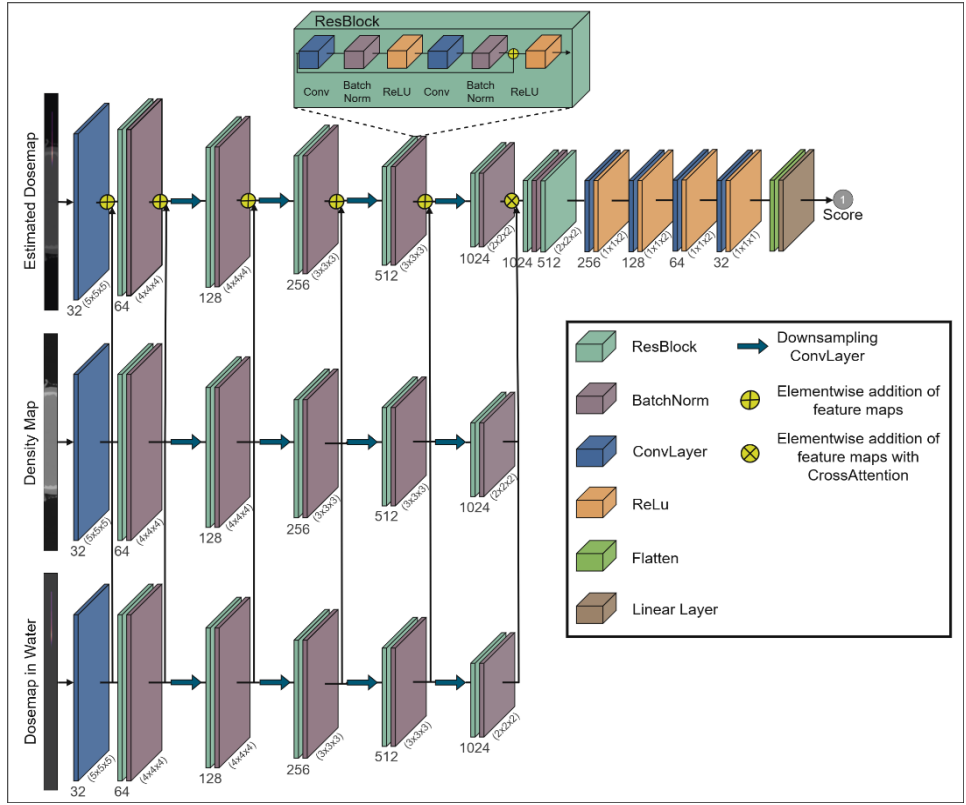

Figure 2: Architecture of the discriminator used for the GAN. The numbers below the boxes represent the respective filter size, while those to the right of the boxes represent the kernel size.

### 2.2.3 Diffusion Model

While a GAN model generates a dose distribution in a single step, Diffusion Models employ an iterative process, which is depicted in Supplementary Figure S5. A noise $\varepsilon$ is incrementally added to the dose distributions in multiple steps $t$ until they become a pure noise distribution $X_T$. The noisy distribution $X_t$ in step $t$ can be calculated by

$$X_t = \sqrt{1 - \beta_t}X_{t-1} + \sqrt{\beta_t}\varepsilon_t. \tag{1}$$

Here, $\beta_t$ encodes the noise intensity and increases linearly with $t$ from $\beta_0 = 1 \times 10^{-4}$ to $\beta_T = 1 \times 10^{-2}$ in $T = 1000$ steps. The noise $\varepsilon_t$ comprises random numbers sampled from a normal distribution.

For the reverse transformation, a model is trained to predict the noise $\varepsilon_t$ added in each step $t$. A slightly modified form of the U-Net described above is employed for this purpose. The modified architecture is depicted in Figure 3. Training is performed using pairs of noisy dose distributions $X_t$ of randomly selected steps $t$, created using equation (1), and the corresponding noise distributions $\varepsilon_t$. Additionally, the model receives the current step number $t$, density matrix, and dose distribution in water as input information. For the time step $t$, a sinusoidal positional encoding[34] is employed. This is integrated via an adaptive group normalization[57]

$$AdaGN(x, t) = \text{Embed}_1(t) * GN(x) + \text{Embed}_2(t), \tag{2}$$

which replaces batch normalization outside the ResBlocks in the U-Net. The time embeddings $\text{Embed}_i(t)$ comprises two linear layers with an embedding dimension of 512 and a Gaussian Error Linear Units (GELU) activation function in between. Within the ResBlocks, batch normalization is replaced by simple group normalization[58].

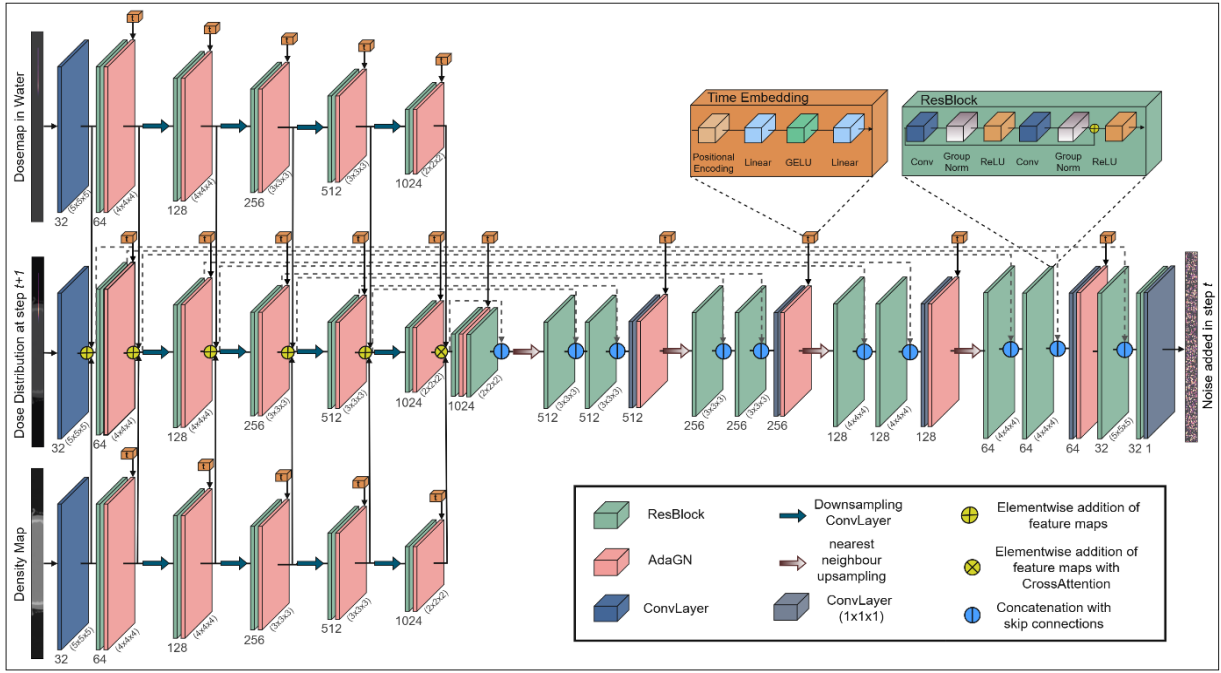

Figure 3: Modified architecture of the U-Net used as a noise estimator in the Diffusion Model. The numbers below the boxes represent the respective filter size, while those to the right of the boxes represent the kernel size.

After training, the Diffusion Model is able to predict the original dose distribution $X_0$ starting with a random noisy image $X_T$ by iteratively subtracting the predicted noise $\varepsilon_\theta$ from the noisy distribution by

$$X_t = \frac{1}{\sqrt{1 - \beta_t}} \left[ X_{t+1} - \frac{\beta_t}{\sqrt{1 - \prod_{i=1}^{t}(1 - \beta_i)}} \varepsilon_\theta \right] + \sigma_t Z, \tag{3}$$

where $\sigma_t$ is zero for the last time step $t = 0$ and $\sigma_t = \sqrt{\beta_t}$ otherwise. $Z$ represents random numbers sampled from a normal distribution and is added to the image to prevent the model from getting stuck in an incorrect or overly deterministic reconstruction.

### 2.2.4 Training Details

All models are trained using the Adam optimizer with an initial learning rate of $10^{-4}$ for the U-Net and Diffusion Model and $10^{-6}$ for the GAN. The learning rate is exponentially reduced during training, with a decay rate of 0.965. Training is conducted on an NVIDIA A100-SXM4

GPU node with 20 GB memory and stops when the validation loss does not decrease over 15 epochs. The batchsize is set to 4.

*2.3 Reference Models*

### 2.3.1 DoTA

The DoTA model is described in detail by Pastor-Serrano et al.[33] and is available on GitHub[59]. The hyperparameter grid search outlined in Pastor-Serrano et al. [33] was performed to adapt the model to our dataset. A larger parameter space was explored than in the original study[33]: The number of transformer layers $N \in \{1, 2, 4\}$, number of convolutional filters $K \in \{8, 12, 16, 20, 24, 28, 32\}$, and number of attention heads $N_h \in \{8, 12, 16, 20, 24\}$. The lowest validation loss is achieved with four transformer layers, 20 convolutional filters, and 20 attention heads. While DoTA was trained for 56 epochs in Pastor-Serrano et al.[33], training on our dataset had not converged after 56 epochs. Therefore, training was extended until the validation loss did not decrease for 15 consecutive epochs, which occurred after 118 epochs.

### 2.3.2 DiffDP

The DiffDP model is used as described in Feng et al.[38]. Instead of segmentation maps, the dose distribution in water is used as an additional input to the CT image. Similar to the previously presented models, the training is terminated after 15 epochs without a decrease in validation loss.

*2.4 Notes on the Model Evaluation*

For the model performance evaluation, only dose depositions within the clinically relevant volume, specifically within the patient, are considered. The region is selected using the density distribution as the area between the first and last voxel along the beam direction where the density exceeds $0.1\,\text{g/cm}^3$. Dose deposited in the air close to the patient, such as between the ears and head, are thus included in the analysis. No dose threshold is applied to the dose distributions.

## 3. Results

*3.1 Analysis of Model Performances*

To assess the robustness and generalizability of the models, the distribution of the relative voxel-wise error between the predicted and true dose distribution of the training, validation and test dataset is compared in Figure 4. This comparison helps to identify potential issues, such as overfitting or systematic biases. The relative error is calculated as the mean of the absolute differences between the dose values of the voxels divided by the maximum dose value of the distribution under investigation. Blue, red and green represent the distribution for the training, validation, and test datasets, respectively.

For all five models analyzed, the distributions exhibit a similar relative shape and comparable range, suggesting that the models effectively learned the underlying relationships within the data rather than memorizing the training data distribution. While the relative error for the U-Net and DoTA leads to a narrow distribution with deviations below $0.5\,\%$, the distributions of the Diffusion Model, GAN, and DiffDP are broader, spanning a range of approximately 1 % for the GAN and 3-5 % for the Diffusion Model and DiffDP.

In addition, Figure 4 illustrates the distribution of the relative error for the unseen patient geometry in yellow. For the U-Net and DoTA, this distribution is shifted and extends considerably wider towards higher deviations (up to about 1 %) than the other datasets. The GAN distribution is approximately 1.5 times broader, with relative errors of up to about 2 %. However, for the Diffusion Model and DiffDP, the unseen geometry distribution roughly matches the broader distributions of the other datasets.

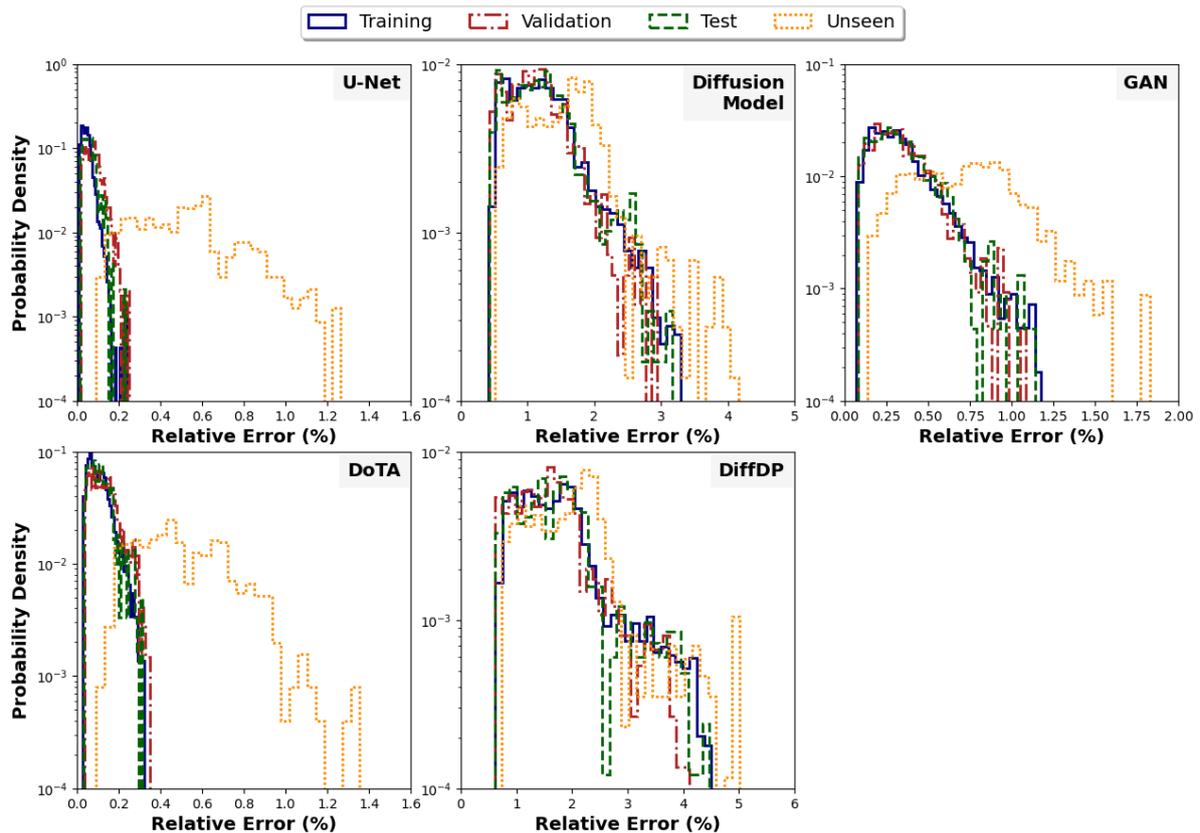

Figure 4: Comparison of the relative error distribution between predicted and true dose distributions for the training (solid blue line), validation (red dash-dotted line), test (green dashed line), and unseen (yellow dotted line) datasets. Each panel represents the error distributions for a specific model (from top left to bottom right: U-Net, Diffusion Model, GAN, DoTA, and DiffDP) for all four datasets.

The mean relative errors for all models of the test and unseen datasets are summarized in Table 1. For a comprehensive evaluation of model performance, the root mean squared error (RMSE) and Gamma passing rate (ΓPR) are also presented for the dose normalized to the maximum dose. The RMSE is more strongly influenced by high-dose regions (at the Bragg peak), which are highly clinically relevant. The Gamma Index[60] is commonly used to compare two dose distributions, considering both dose values and spatial deviations. The ΓPR represents the percentage of voxels that meet predefined criteria for maximum deviation of relative dose and position of data points. In this study, the criteria applied include a maximum dose deviation of 3 % and maximum spatial deviation of 3 mm (3 %/3 mm), as well as a stricter criterion of a maximum dose deviation of 1 % and maximum spatial deviation of 1 mm (1 %/1 mm).

The lowest errors on the test dataset are achieved by the U-Net. However, on the unseen dataset, DoTA achieves slightly lower RMSE than the U-Net. The relative errors of all models for the test and unseen datasets are roughly in the same range and increase slightly by approximately 0.5 % for the unseen dataset. Conversely, the RMSE shows an increase of about an order of magnitude for the unseen dataset compared to the test dataset, indicating weaknesses in the models' ability to accurately capture the Bragg peak for the unseen dataset (a detailed analyzation of the model's performance in the Bragg peak region is presented in Figure 5). The (3 %/3 mm) ΓPR is above 99 % for all models on both the test and unseen datasets, with a slight decrease of approximately 0.1–0.2 % on the unseen dataset. The U-Net, GAN, and DoTA also achieve ΓPRs of approximately 99 % for the stricter (1 %/1 mm) criterion, while the Diffusion Model reaches only around 92 % and DiffDP achieves about 95 %. For the (1 %/1 mm) criterion, the ΓPR for all models decreases slightly, by less than 1 %, on the unseen dataset.

Table 1: Mean and standard deviation of the relative error, RMSE of the relative dose, and ΓPR between the predicted and true dose distributions for the test and unseen datasets across all investigated models.

| | Relative Error (%) | RMSE ($10^{-3}$) | ΓPR (3 %/3 mm) (%) | ΓPR (1 %/1 mm) (%) |
|---|---|---|---|---|
| **Test Dataset** | | | | |
| **U-Net** | 0.05 ± 0.03 | 0.99 ± 0.47 | 99.86 ± 0.11 | 99.70 ± 0.17 |
| **Diffusion Model** | 1.23 ± 0.54 | 6.92 ± 1.21 | 99.88 ± 0.10 | 92.31 ± 5.65 |
| **GAN** | 0.34 ± 0.18 | 2.68 ± 0.73 | 99.85 ± 0.10 | 99.20 ± 0.86 |
| **DoTA** | 0.11 ± 0.06 | 1.55 ± 0.43 | 99.81 ± 0.12 | 99.65 ± 0.23 |
| **DiffDP** | 1.74 ± 0.77 | 7.02 ± 0.52 | 99.91 ± 0.07 | 95.19 ± 3.19 |
| **Unseen Dataset** | | | | |
| **U-Net** | 0.51 ± 0.23 | 20.21 ± 4.68 | 99.62 ± 0.21 | 99.07 ± 0.34 |
| **Diffusion Model** | 1.56 ± 0.66 | 23.22 ± 4.28 | 99.75 ± 0.19 | 92.78 ± 5.64 |
| **GAN** | 0.72 ± 0.31 | 20.13 ± 5.26 | 99.64 ± 0.21 | 98.75 ± 0.67 |
| **DoTA** | 0.51 ± 0.22 | 18.44 ± 5.61 | 99.61 ± 0.23 | 99.06 ± 0.39 |
| **DiffDP** | 2.04 ± 0.85 | 22.85 ± 4.31 | 99.72 ± 0.20 | 95.25 ± 2.91 |

Four parameters characterizing the Bragg peak were compared to enable a detailed evaluation of the models' performance in the clinically relevant Bragg peak region, shown in Figure 5. The distributions of the deviations between predicted and ground truth values of the parameters are depicted using turquoise and yellow for the test and unseen datasets, respectively. The data are presented as a violin plot (using the seaborn library with default parameter[61]), where the distribution of the deviations is depicted as a kernel density estimate. The middle-dashed lines indicate the medians of the distributions, while the dotted lines are the quartiles. The distributions were cropped for better clarity; the same plot without cropping is provided in Supplementary Figure S6.

As expected, all models exhibit a broader distribution for the unseen dataset than the test dataset with all parameters. Figure 5(a) shows the 90 % distal fall-off position $\text{Depth}(D_{90})$ as a measure of the range of the carbon ions. While a narrow-centered distribution can be observed for the test dataset for the U-Net, the distribution for the unseen dataset is shifted towards higher values, which corresponds to a higher range. In contrast, the GAN and DiffDP tend to predict shorter ranges for the test dataset, while the distributions for the unseen dataset are centered. The Diffusion Model and DoTA show centered distributions of similar width for both datasets.

The position of the Bragg peak perpendicular to the beam direction is predicted with relatively high accuracy by all models, as illustrated in (b). While the deviation for the test dataset is zero voxels in most cases, the Bragg peak for the unseen dataset is often shifted by 1 or 1.5 voxels.

Figure 5(c) examines the relative deviation of the predicted maximum dose value from the real maximum dose value. Systematic negative deviations are observed for all models, indicating an underpredicted Bragg peak height. These deviations are particularly pronounced for U-Net, GAN, and DoTA on the unseen dataset.

Deviations in the width of the Bragg peak in beam direction, represented by the full width at half maximum (FWHM), are analyzed in Figure 5(d). While narrow distributions centered around zero voxels are observed for all models on the test dataset, the unseen dataset distributions are broader and shifted toward negative values for all models. This indicates that the models tend to predict sharper Bragg peaks.

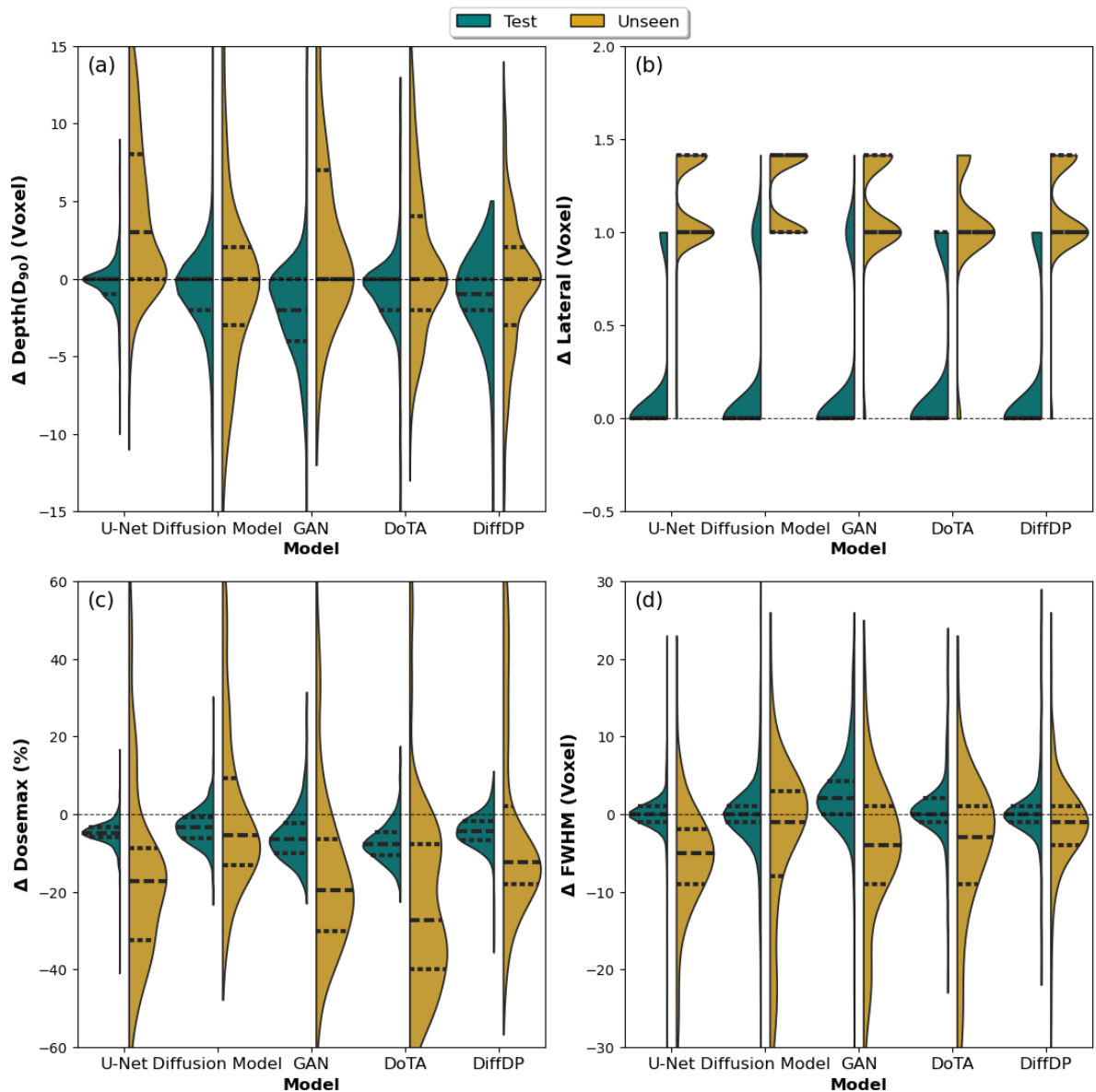

Figure 5: Comparison of the deviations between Bragg peak characteristics predicted by the investigated models and the corresponding true values for the test (turquoise) and the unseen (yellow) datasets. The deviations are shown for the 90% distal fall-of position $\text{Depth}(D_{90})$ (a), Bragg peak position perpendicular to the beam direction (b), maximum dose value (c), and FWHM (d). The colored areas of the violin plot represent the kernel density estimates of the deviations, the dashed line denotes the median, and the dotted lines are the quartiles. The distributions were truncated for better clarity. A plot of the full range is provided in Supplementary Figure S6. The black dashed line at zero serves as a guide for the eye.

In addition to overall performance, assessing the performance on partial datasets with the same characteristics is crucial. Figure 6 shows the absolute deviation of the 90% distal fall-of position (top) and maximum dose value (bottom) of the test data for the different patient geometries (left) and carbon ion energies (middle) for all models investigated. The plot on the right shows the energy dependence of the deviations for the unseen dataset. The deviations are represented as blue crosses for U-Net, pink squares for GAN, red circles for the Diffusion

Model, green triangles for DoTA, and yellow diamonds for DiffDP. The symbols indicate the mean value, while the error bars show the standard deviation.

Considering the deviations in the 90 % distal fall-of position, all models exhibit relatively consistent performance across the different patients, with variations within a range of a few voxels. Similarly, the deviations of the maximum dose value vary only by a few percent between the patients for all models.

The deviations of the maximum dose value and 90 % distal fall-of position show no systematic dependence on carbon ion energy for the test dataset. For the unseen dataset, the deviations of the 90 % distal fall-of position increase slightly for all models, reaching a maximum at around 2100 MeV (175 MeV/u) before they quickly decrease. For the deviation of the dose maximum, a minimum can be seen at the same point.

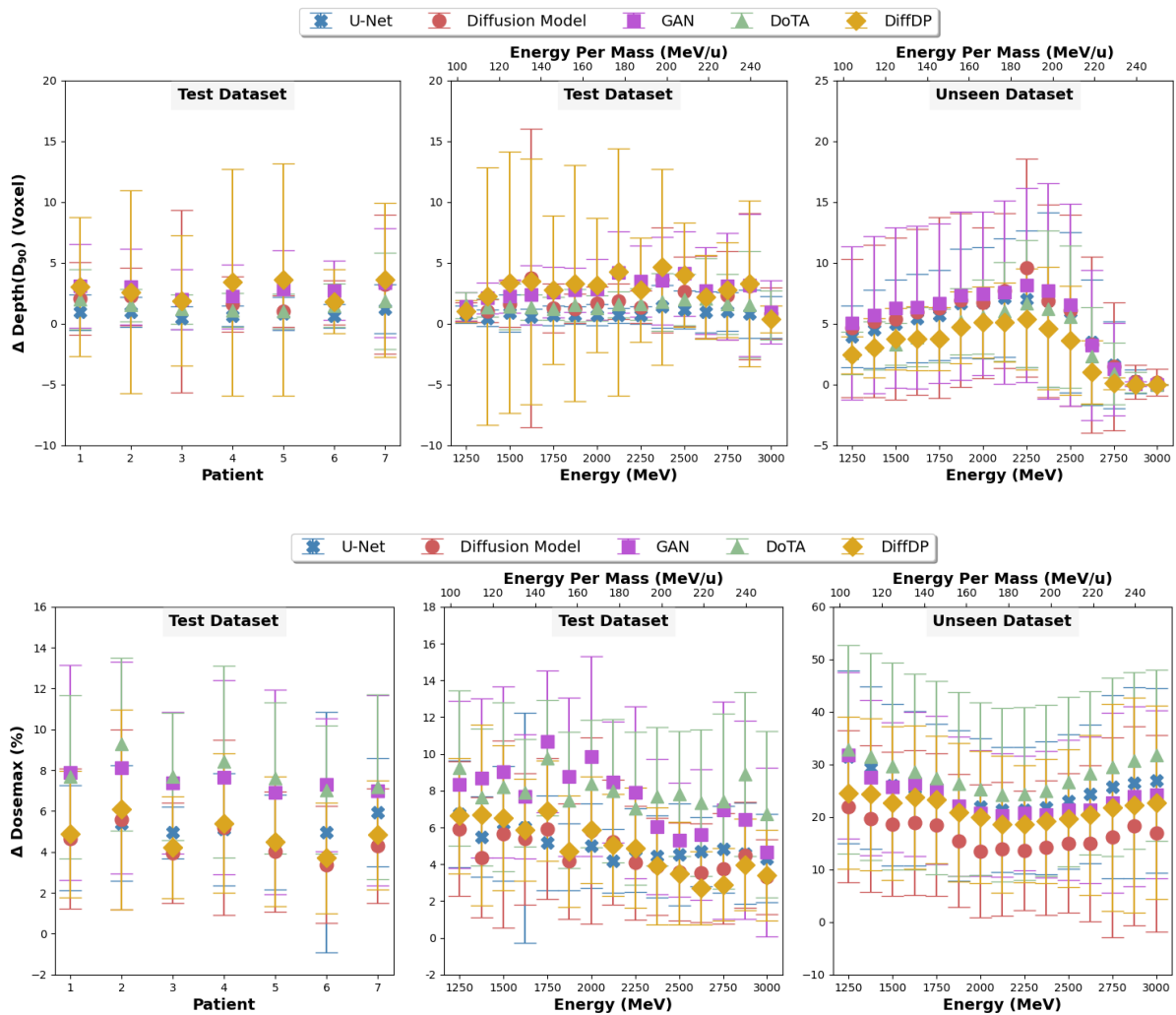

Figure 6: Mean of the absolute deviations of the 90% distal fall-off position (top) and maximum dose value (bottom) between the predicted and true dose distributions as a function of the patient and carbon ion energy. The deviations of the U-Net prediction are shown in blue crosses, those of the Diffusion Model in red circles, those of the GAN in pink squares, those of the DoTA in green triangles, and those of the DiffDP in yellow diamonds. The left panels show the dependence of the deviations on the patient for the test dataset, the middle panels indicate the energy dependence for the test dataset and the right panels demonstrate the energy dependence for the unseen dataset. The error bars represent the standard deviations.

### 3.2 Predicted Dose Distribution Examples

An investigation of the worst performing examples is useful, as it highlights systematic weaknesses and specific challenges for the models. Figure 7 compares the depth-dose curves predicted by all the analyzed models for the samples from the test dataset with the highest relative error. Similarly, Figure 8 displays the worst performing samples from the unseen dataset. Each panel illustrates the predicted depth-dose curves of the same sample for all models. The predictions of the U-Net are shown in blue, the Diffusion Model in red, the GAN in pink, the DoTA in green, and the DiffDP in yellow. The true depth-dose curve is displayed in black. The model for which this sample exhibits the highest relative error is indicated by the name in the top left corner, while the corresponding relative error is indicated in the top right corner. The carbon ion energy and patient number of the sample are also provided at the top of the panel. The depth-dose curves were obtained by extracting the dose values along that axis parallel to the beam, including the maximum dose value of the distribution (Geant4 ensures that the source position is not located on a voxel boundary.). Additionally, the density distributions along the beam axis are shown as dashed lines corresponding to the right y-axes. The corresponding 2D distributions of the worst-performing examples of each model are shown in Supplementary Figure S7 for the test dataset and Supplementary Figure S8 for the unseen dataset.

The worst-performing test samples often involve high energy combinations and untypical geometric situations. These include cases with many density variations due to anatomical structures, such as the nasal or ear regions or Bragg peak locations close to or within the skull bone. In the worst sample of the U-Net, the model appears to ignore the density increase upon entering the head. While the Bragg peak position is correctly predicted, its magnitude is overestimated. Although the general shape of the depth-dose curve is captured for the worst predictions of the Diffusion Model, GAN, DoTA, and DiffDP, the dose values are underestimated. Additionally, DiffDP predictions exhibit considerable noise in many cases, particularly in its worst-case example.

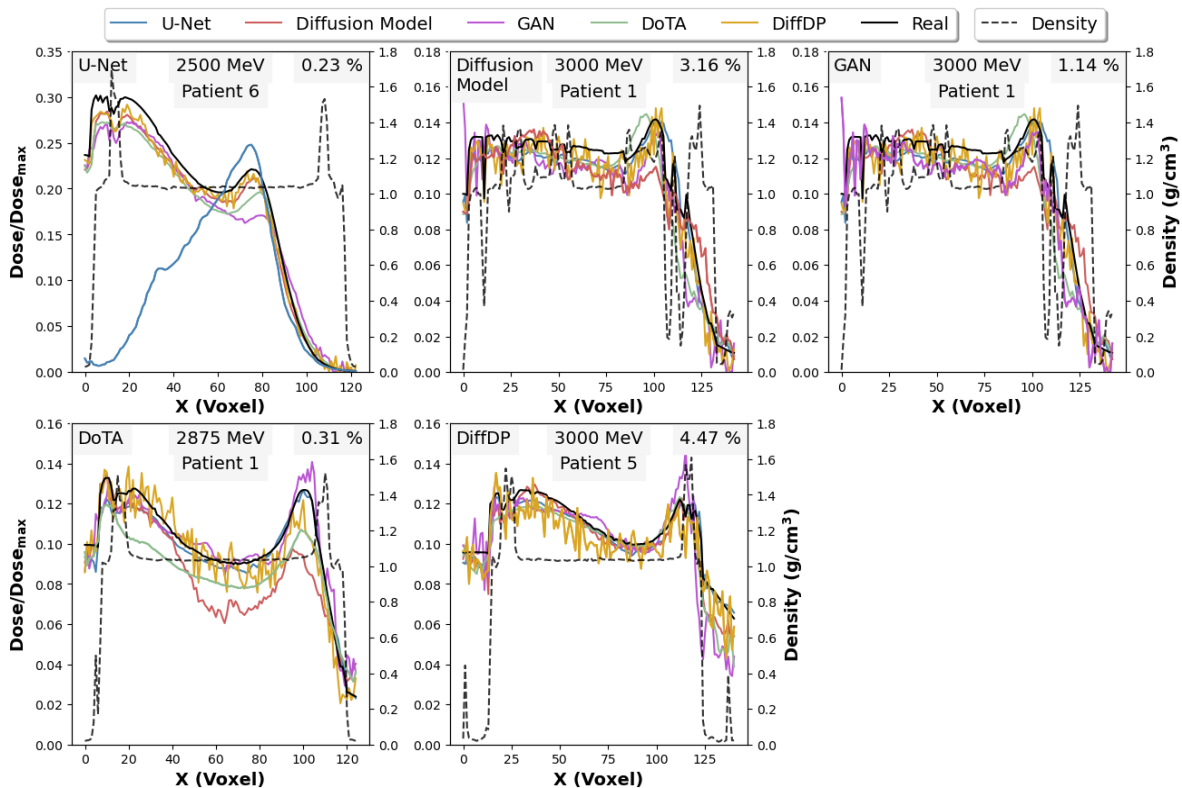

Figure 7: Depth-dose curves (dose values along the axis parallel to the beam that contains the dose maximum of the respective dose distribution) of the investigated models for the predicted dose distribution with the highest relative error from the test dataset. Each panel shows the predicted depth-dose curves of all models for the same sample. The U-Net prediction is represented in blue, the Diffusion Model in red, GAN in pink, DoTA in green, and DiffDP in yellow. The true dose distribution is shown in black, while the density along the beam direction at the source position is represented by a dashed black line. The panels, from the top left to the bottom right, show the examples with the highest error for U-Net, Diffusion Model, GAN, DoTA, and DiffDP prediction, respectively. The carbon ion energies, patient numbers, and relative errors of the model under consideration are indicated for each sample at the top of each panel.

The worst examples for the unseen dataset also involve cases where the Bragg peak is near the skull bone. Significant over- or underestimations of the dose can be observed in all instances. Remarkably, the models often predict a similar shape and height of the dose curve, even in cases where this deviates from the true dose distribution.

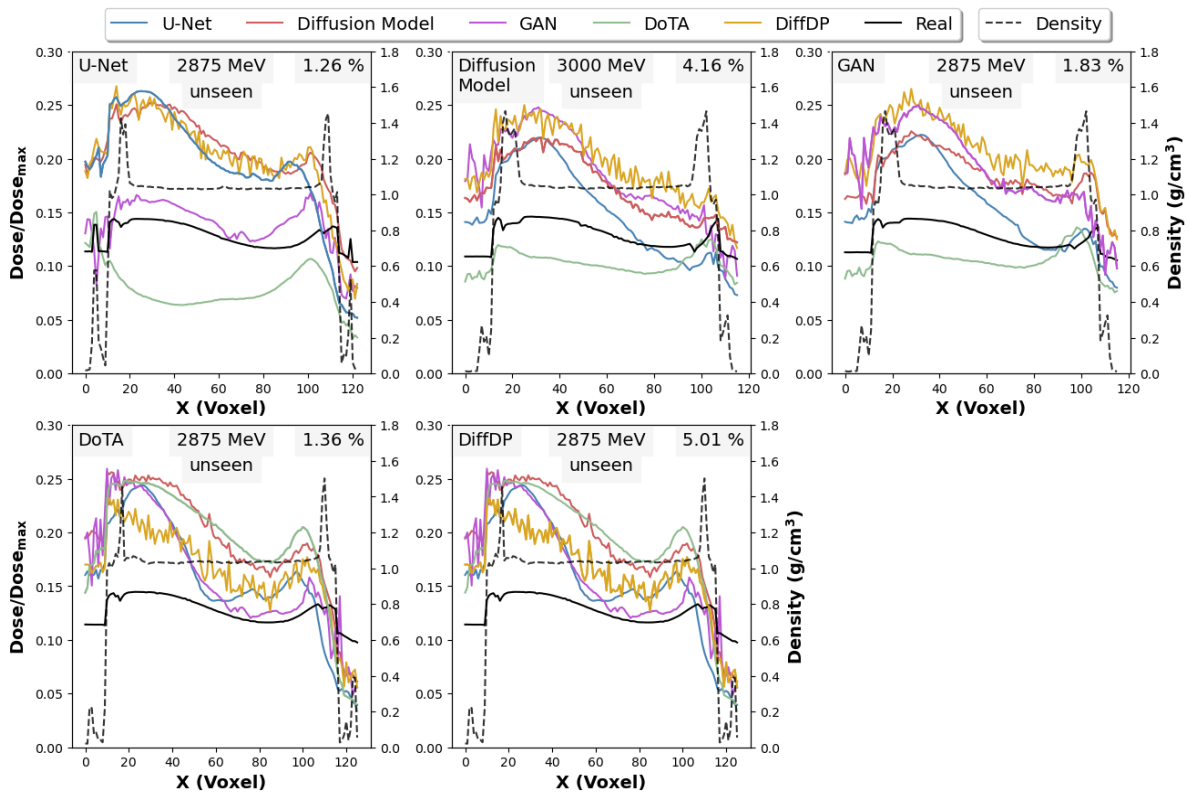

Figure 8: Depth-dose curves (dose values along the axis parallel to the beam that contains the dose maximum of the respective dose distribution) of the investigated models for the predicted dose distribution with the highest relative error from the unseen dataset. Each panel shows the predicted depth-dose curves of all models for the same sample. The U-Net prediction is represented in blue, the Diffusion Model in red, GAN in pink, DoTA in green, and DiffDP in yellow. The true dose distribution is shown in black, while the dashed black line represents the density along the beam direction at the source position. The panels, from the top left to the bottom right, are the examples with the highest error for U-Net, Diffusion Model, GAN, DoTA, and DiffDP prediction, respectively. The carbon ion energies, patient numbers and relative errors of the model under consideration are indicated for each sample at the top of each panel.

*3.3 Dose Calculation Time*

Table 2 shows the average times for calculating a single dose distribution using the MC simulation and all models investigated. The simulation time represents the duration of simulating one sample, which is averaged over all samples in the dataset. It is multiplied by the number of nodes used in parallel processing. To determine the computation time for the dose prediction models, the prediction for the test dataset was repeated 10 times, and the mean time for predicting a single dose distribution was determined.

Table 2: Overview of dose calculation times for generating a single dose distribution for all investigated models and the MC simulation. The MC simulation time represents the average time required to simulate a single dose distribution during dataset creation, multiplied by the number of central processing unit (CPU) cores used (25 CPU cores were utilized). The times for the dose prediction models represent the mean and standard deviation of a 10-fold repetition of the dose predictions for the test dataset. The batch size was set to 4, as during the training.

|  | Dose calculation time (s) | |
|---|---|---|
|  | 1 GPU | 1 CPU |
| **U-Net** | $(11.3 \pm 0.2) \times 10^{-3}$ | $3.95 \pm 0.14$ |
| **Diffusion Model** | $15.826 \pm 0.005$ | $(4.7 \pm 0.2) \times 10^3$ |
| **GAN** | $(11.2 \pm 0.2) \times 10^{-3}$ | $4.1 \pm 0.2$ |
| **DoTA** | $(3.022 \pm 0.003) \times 10^{-3}$ | $0.368 \pm 0.002$ |
| **DiffDP** | $6.931 \pm 0.004$ | $(1.28 \pm 0.04) \times 10^3$ |
| **MC Simulation** | - | $2.3 \times 10^5$ |

## 4. Discussion

One key technical challenge in treatment planning for carbon ion therapy is finding a balance between the accuracy of dose calculation algorithms and their computation times. Numerous studies from other radiotherapy modalities have demonstrated that this can be achieved using ML models[25,28,33,35,43]. The results presented here indicate that this approach can also be applied to carbon ion therapy.

All presented models accelerate dose calculation by several orders of magnitude compared to MC simulations (Table 2). A dose distribution can be computed within milliseconds to seconds when using a GPU. These computation times are comparable to previously reported prediction times for dose distributions with other radiation modalities[25,28,33,43].

The dose prediction accuracy was analyzed using two datasets: A test dataset comprising unseen samples from previously seen patients and an unseen dataset with a completely new patient. All models achieve an average ΓPR of over 99 % for the (3 %/3 mm) criterion on both the test and unseen datasets, as shown in Table 1. A dose distribution is often accepted if at least 95 % of the voxels fulfill the (3 %/3 mm) criterion. All predicted dose distributions meet this requirement. Moreover, the stricter (1 %/1 mm) criterion is fulfilled in over 95 % of the voxels for all models except the Diffusion Model, which achieves an average ΓPR of approximately 92 % under this criterion. On both datasets, mean relative errors of less than 2 % are achieved for all models, only DiffDP is just above 2 % for the unseen dataset.

A crucial clinical consideration is accurate dose modeling in the Bragg peak region, which is examined in detail in Figure 5. For the test dataset, the Bragg peak position is predicted correctly in almost all cases in the dimensions perpendicular to the beam axis. The predicted range, represented by the 90 % distal fall-off position, deviates by less than five voxels from the true range for most dose distributions in the test dataset. The same can be observed for the Bragg peak width. However, both distributions exhibit significant outliers. All models systematically underestimate the height of the Bragg Peak by approximately 5 % for the test dataset and 15 %–30 % for the unseen dataset. As shown in Figure 6, this underestimation seems independent of energy and patient.

For the unseen dataset, the prediction of the carbon ion range becomes less accurate: The distribution of deviations in the 90 % distal fall-off position is broader and has more outliers. Notably, the Bragg peak position perpendicular to the beam direction is missed by 1 to 1.5 voxels in all cases, which was rarely observed in the test dataset. The underestimation of the maximum dose value is pronounced, and the distributions, particularly for U-Net, GAN, and DoTA, spread toward negative values.

The CT images in the dataset use variant image resolutions, while previous studies included only patients whose CT images were acquired with identical resolutions. Resampling the voxel size using the nearest-neighbor interpolation method may create softer or harder edges. The latter causes abrupt changes in density or dose. Such discontinuities are particularly expected in regions with high density or dose gradients, such as the Bragg peak region or around the skull. Difficulties in dose prediction in regions with large density gradients have been documented[28], independent of resampling uncertainties. Resampling uncertainties are expected to exacerbate this effect. Most of the worst-performing examples in Figure 7 and Figure 8 are cases where high dose and density gradients coincide or where many density fluctuations are present.

When comparing the models with each other, U-Net and DoTA exhibit the smallest deviations (Table 1 and Figure 1). However, the Diffusion Model and DiffDP show a smaller performance gap between the unseen and the other datasets, indicating improved generalizability of these models to different CT image resolutions. Figure 5 shows that this is due to the reduced underestimation of the dose values and Bragg peak width.

All models exhibit relatively similar behavior in the Bragg Peak region (Figure 5 and Figure 6). The DoTA model, which lacks information about the dose distribution in water, shows no notable differences, except for a slightly stronger underestimation of the Bragg Peak height than the other models.

The results presented here demonstrate that the generalization of dose prediction models to different CT images with varying resolutions is feasible. The observed performance gap is expected to be reduced by incorporating an increased variety of CT images into the training process. In general, a larger training dataset with more variance in the data is expected to increase the accuracy of the models. In particular, increasing the proportion of more complex geometries as in the worst-performing examples in the training dataset should lead to higher accuracy for such cases.

In future work, it would also be valuable to investigate the behavior of the models for patients with geometric peculiarities, such as dental fillings, and to extend the study to other anatomical regions. Additionally, including different beam sizes and properties beyond beam energy remains vital for further investigation. It would also be of interest to explore whether the approach, so far applied only to pencil beams, can be extended to the calculation of complete radiation fields. To integrate dose prediction algorithms into clinical practice, also a mechanism for identifying cases with a high probability of incorrect predictions is crucial. This requires further research into applying explainability methods in dose prediction.

Furthermore, treatment planning in carbon ion therapy necessitates incorporating the radiation RBE[6,62]. This requires concepts beyond conventional dosimetry, concepts from micro- and nanodosimetry[63–65]. The detailed characterization of energy deposition at these scales is significantly more computationally expensive than a calculation of a voxel-based dose distribution. Exploring whether ML models can consolidate accuracy and computational efficiency in these cases is imperative.

## 5. Conclusion

This study advances the potential of ML models for dose prediction by investigating their generalization to carbon ion therapy and different CT image resolutions. The results demonstrate that fast and accurate dose prediction using ML models is in principle feasible for carbon ion therapy. Previously introduced models, such as DoTA and DiffDP, can be adapted to this radiotherapy modality and dataset with minimal modifications. Additionally, the study highlights that computationally intensive models, such as GANs or Diffusion Models, do not offer substantial performance advantages over less resource-intensive models, such as the U-Net. Beyond evaluating these models for additional data variations, such as different beam properties or patients with artifacts, future research should improve model interpretability and provide uncertainty estimates for predictions.

## 6. Acknowledgments

The authors thank the team of the High-Performance Computing Cluster at Physikalisch-Technische Bundesanstalt (PTB) for their ongoing support and Mr. Ali Kamal for running some of the simulations that assisted in creating the dataset. This work was supported by the "Metrology for Artificial Intelligence in Medicine (M4AIM)" program, funded by the German Federal Ministry of Economic Affairs and Climate Action in the frame of the QI-Digital Initiative.

## 7. Data Availability Statement

The data supporting the findings of this study are openly available at the following URL/DOI: https://gitlab1.ptb.de/MiriamSchwarze/mlmodelsforcarboniontherapy.